\documentclass[a4paper]{article}
\usepackage{icrc2013}
\usepackage[english]{babel}

\title{Suprathermal particle addition to solar wind pressure: possible influence on magnetospheric transmissivity of low energy cosmic rays?}
\shorttitle{Suprathermal solar wind pressure}

\authors{P. Bobik$^{5}$, M. J. Boschini$^{1,6}$, C. Consolandi$^{1}$, S. Della Torre$^{1}$, M. Gervasi$^{1,4}$, D. Grandi$^{1}$, K. Kudela$^{5}$, G. La Vacca$^{1,4}$, M. Mallamaci$^{1}$, S. Pensotti$^{1,4}$, P.G. Rancoita$^{1}$, D. Rozza$^{1,2,3}$, M. Tacconi$^{1,4}$.}
\afiliations
{
$^1$ \textit{INFN Sezione di Milano Bicocca, I-20126 Milano, Italy}\\
$^2$ \textit{Universit\'a dell'Insubria, I-22100 Como, Italy}\\
$^3$ \textit{European Organization for Nuclear Research, CERN, CH-1211 Geneva 23, Switzerland}\\
$^4$ \textit{Universit\'a di Milano Bicocca, I-20126 Milano, Italy}\\
$^5$ \textit{IEP SAS Kosice, Slovakia}\\
$^6$ \textit{CINECA} (Italy)\\
}
\email{stefano.dellatorre@mib.infn.it}

\abstract{Energetic (suprathermal) solar particles, accelerated in the interplanetary medium, contribute to the solar wind pressure, in particular during high solar activity periods. We estimated the effect of the increase of solar wind pressure due to suprathermal particles on magnetospheric transmissivity of galactic cosmic rays in the case of one recent solar event. }


\begin{document}

\maketitle

\section{Introduction}
Numerical tracing of charged particle trajectory in the geomagnetic field models has been extensively used for long time as a tool for understanding the magnetospheric transmissivity of low energy cosmic rays (rev. e.g. in \cite{Shea}). In order to estimate the transmissivity the geomagnetic field model adopted is also important (rev. e.g. in \cite{Desorgher}). In recent decades several geomagnetic field models including also the external current systems have been developed. Strong geomagnetic disturbances lead to changes in transmissivity parameters (e.g. \cite{Kudela, Tyasto}) and the results of transmissivity depend on the model used. In the external field models (TS89, TS96, TS02, TS05, the complete references in \cite{Tsyganenko}) one of the input parameters is the solar wind dynamic pressure.
After solar flares, during CME transit, low energy particles are accelerated in the interplanetary space close to the Earth, especially when crossing  shock waves. These energetic particles can contribute to the total particle pressure with the effect to change the magnetopause and bow shock position. Here using the method of computations described in \cite{ICRC-geomag} we estimate the effect of the suprathermal particle contribution to the total dynamic pressure of the solar wind and use this result as input to the geomagnetic field model.

\section{Data}
Solar wind pressure data, measured by several space probes, are available at the web site in ref. \cite{omniweb}. These data are obtained compiling measurements of the lowest energetic components of the solar emission. However, especially during solar events, CMEs and associated shock motion in interplanetary space, particles are accelerated at the discontinuities. These energetic particles can arrive to the vicinity of Earth earlier than CME impact and can contribute to the dynamic pressure. The time sequence of the impact of the several components at the Earth could be important for the interaction with the magnetosphere. For selected events we use data from EPAM/ACE instrument \cite{ACE} of ions at different energies in the interval 56 -- 3020 keV and compute the suprathermal ion pressure.
The EPAM/ACE instrument has different detectors, but we consider the LEMS120 detector (Low-Energy Magnetic Spectrometer). For each of the eight channels ($P_1-P_8$) we have the average energy ($E_i$) and the energy width ($\Delta E_i$) as described in \cite{Gold}. For characteristic speed the formula $v_i = (E_i/m)^{1/2}$ is used. Here $m$ is an effective mass including a small contribution of helium $m=(0.955+4*0.045)*m_p$ and $m_p$ is the proton rest mass. The ratio of number densities, consistently with omniweb-NASA \cite{omniweb}, is taken to be $0.045=n_{He}/n_p$. For each of the 8 channels ($P_i$) we use the 5 minute-resolution flux and calculate the corresponding pressure of suprathermal particles as $p_i(t) = 4 \pi \Phi(t) \Delta E_i m v_i$, where $\Phi(t)$ is the time depending particle flux. The suprathermal contribution ($p_{st}$) to the solar wind pressure is therefore the sum of all the measured channels: $p_{st} = p_1(t)+p_2(t)+...+p_8(t)$ for each time $t$. Strong variations of the two pressures and of their ratio are observed during the disturbed interplanetary events. Out of the events with CMEs and accelerated particles seen by EPAM/ACE, we selected and plotted in Figure 1 the one occurred during January 2012.

\begin{figure}[!t]
  \centering
  \includegraphics[width=0.5\textwidth]{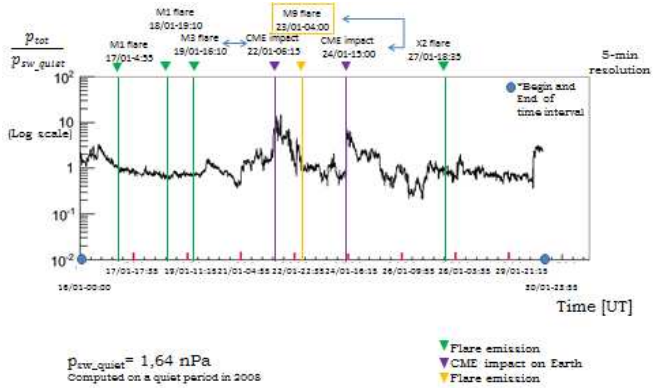}
  \caption{Time profile of the ratio of the total pressure (solar wind + suprathermal particles) for the period 16 - 30/01/2012. The timing of flares and CMEs arrival to the Earth is also shown.}
  \label{SWpressure1}
 \end{figure}

\begin{figure}[!t]
  \centering
  \includegraphics[width=0.5\textwidth]{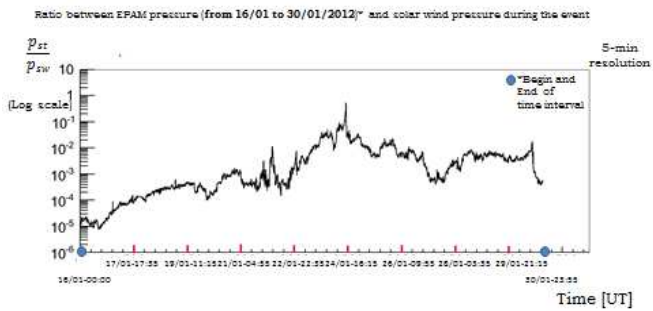}
  \caption{Ratio of the pressure due to suprathermal particles to the solar wind pressure. The maximum contribution of suprathermal particles to the pressure ($p_{st} / p_{sw} \sim 0.6$) occurs on the day 24/01/2012 at $\sim$ 14:40 UT, about 20 minutes before CME impact.}
  \label{SWpressure2}
 \end{figure}

\section{Transmissivity computations}
Using the GCR energy spectra (see \cite{AMS}) we computed the transmissivity through the magnetosphere to the site of a satellite low altitude orbit.  We compared the effect with and without the addition of the suprathermal particle components. We adopted an increase of the solar wind pressure ($p_{dyn}$) of a factor $\sim 1.6$ for two regions: where there is the maximum vertical cut-off rigidity (near the equator) and  where is the maximum geomagnetic latitude reached by ISS. The computations are done assuming an ideal detector with $2 \pi$ angular acceptance (see Figs 3 and 4). We used two models of geomagnetic field and two different approaches for the magnetopause position. Using the model TS96 the estimated effect on the total flux is 0.85 \% at high latitudes and 0.34 \% at low latitude.

\begin{figure}[!t]
  \centering
  \includegraphics[width=0.5\textwidth]{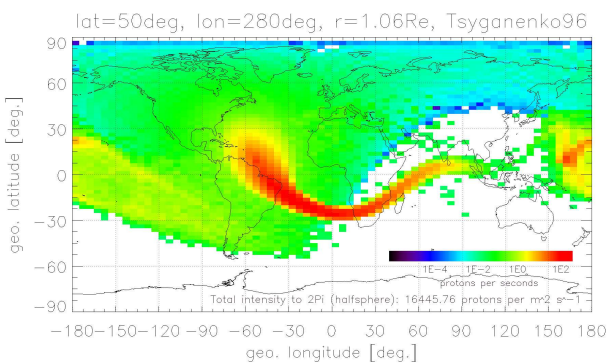}
   \caption{Contributions to the expected flux of GCR measured by an ideal detector with omnidirectional $2 \pi$ acceptance at the position 50N, 280E, 1.06 $R_e$ from various asymptotic directions. Model Tsyganenko 96 is used and nominal $p_{dyn}$ from solar wind included for the event in Fig.1.}
   \label{Mag-transmitt1}
 \end{figure}

 \begin{figure}[!t]
  \centering
  \includegraphics[width=0.5\textwidth]{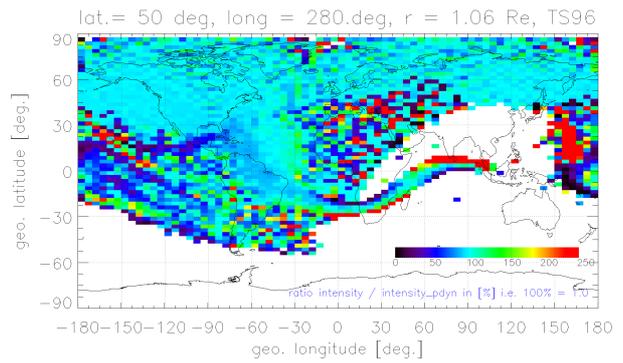}
   \caption{Differences (ratios) between the data plotted in Fig.3 and the distribution of fluxes when $p_{dyn}$ is multiplied by 1.6 corresponding to peak in Fig.2.}
   \label{Mag-transmitt2}
 \end{figure}

\section{Concluding remarks}
A first estimate of the effect of the suprathermal particle flux contribution to the solar wind pressure on the transmissivity of the magnetosphere to GCR indicates a rather small effect to omnidirectional flux observed at low altitude. Its fine features can be obtained using higher statistics than here. This effect may be of some relevance for detectors with very high geometrical factors and fine characteristics of angular acceptance when the orientation of the detector is precisely known.

\textbf{Acknowledgement}. This work is supported by Agenzia Spaziale Italiana under contract ASI-INFN I/002/13/0, Progetto AMS - Missione scientifica ed analisi dati. KK wishes to acknowledge VEGA grant agency project 2/0040/13 for support. Finally, the authors acknowledge the use of
NASA/GSFC’s Space Physics Data Facility’s OMNIWeb service, and OMNI data.

\end{document}